# Worlds beyond our own

**A Midsummer Night's Daydream!**

Lying wide-awake in a cool midsummer night on the house terrace, a pastime that has these days become rare, some daydreamer amongst us looking up at the star-studded sky might have wondered if there is someone (an intelligent being!) up there in an alien world around one of the far-distant stars, right now looking down (looking up from its own point of view!) at our world and in turn wondering if there is some alien (us!) up there right now looking down… Or are we instead alone in our Milky Way galaxy or even in the whole Universe?  May be we need to pose a more pertinent question first – Do other Earth-like worlds exist out there in the space yonder that could support life? In fact a still simpler query would be – Are there other planetary systems like our Solar system that abound in our Galaxy? A mere quarter century back such questions would have been largely academic in nature with no definite answer. But in the past two decades or so a large number of extra-solar planetary systems have been found with some of them having apparently quite earth-like characteristics. Still there is no definite answer to the question about the presence of even the simplest forms of life elsewhere, forgetting the question of an intelligent life. We have only one example of the formation of life, which is on our Earth here, and we do not yet know what it all takes to start the formation of life in the first place. But once the simplest forms of life get triggered, higher forms of life perhaps may not be too difficult to come about, though they might look so vastly different from us that we may not even recognize them as living beings even if they happen to sprung up right in our own backyard.

**The First Discovery of an Exoplanet**

With not even an inkling about the existence of other worlds beyond our own, the ancient doctrine that Earth is the centre of everything and is thus unique, remained almost undisputed throughout the human history till it got challenged by the Copernican heliocentrism in the sixteenth century. Giordano Bruno was among the first ones who went beyond the Copernican model and put forward the view that the fixed stars are similar to the Sun and likewise accompanied by planets. According to him all these planets constituted an infinite number of inhabited worlds, a philosophical position known as cosmic pluralism. For his 'heresies' Bruno was burnt on stake in 1600. Though the search for such planets, now popularly termed as exoplanets, started in the late nineteenth century but the first breakthrough came only a century later when there were reports of the discovery of an extra-solar planet around a pulsar. After an initial false start, it was in the year 1992 that our belief in the uniqueness of our solar system in the universe was finally shattered by the first confirmed detection[1] of two extra-solar planets around a 6.22 millisecond pulsar PSR 1257+12. This was soon followed by discoveries of Jupiter-sized gas giants around Sun-like stars 51 Pegasi[2], 47 UMa[3] and 70 Vir[4]. Currently there are more than 1800 exoplanets that are already discovered and the number is exponentially increasing with the improving observing facilities and instrumentation capabilities.

**The methods for exoplanet detection**

The zeal for hunting other worlds has galvanized astronomers into developing several techniques in order to search for planetary systems beyond our own Solar System. These techniques include the traditional transiting method which measures the drop in starlight when a planet transits across the disk of the star and the astrometry method which measures the reflex motion of the star based on series of images over long times. Microlensing method, which measures an apparent change in the starlight of a distant star due to the gravitational bending by the presence of an intervening star having a companion, has been capable of detecting a few exoplanets. However, radial velocity (RV) method for detecting exoplanets has turned out to be one of the most reliable methods till date. A brief overview of each of the methods for detection of exoplanets is given below.

*Pulsar Timing Method*

The discovery of an exoplanet came from observations of slight modulations of the pulse period of a pulsar. Pulsars are compact neutron stars that may be left behind in the final stages of a sufficiently massive star's life cycle. These compact stars have masses slightly more than that of the Sun contained within a compact sphere of about 10 km radius. The density is so high that a tea-spoonful material of a neutron star will have a mass of about 500 million tons. These compact stars rotate about their axes so fast that it takes only about a second to complete a rotation, though there are pulsars which complete their rotations in almost a milli-second. A pulsar gives out a constant source of radio emission along its magnetic axis which, in general, may be inclined to the rotation axis and an observer sees a pulse of emission whenever the magnetic axis during the star's rotation cuts through the observer's line of sight. These pulse periods are very precise and any variations in them can be detected to accuracies as high as a few femto ($10^{-15}$) seconds. It is the study of such variations that led to the discovery of the very first extra-solar planet. Any alteration in this period may indicate the presence of a companion. It is not just the planet which goes around its host star, but the star-planet system that goes around their barycentre (common centre of mass) with the same revolution period. But the radius of the star's orbit is smaller as compared to that of the planet in the ratio of their masses. As the star moves in an orbit, it causes tiny variations in its observed pulse period. Radio telescopes can easily pick up these tiny modulations in the pulse period of the pulsar, leading thereby to the detection of a companion to the rotating neutron star.

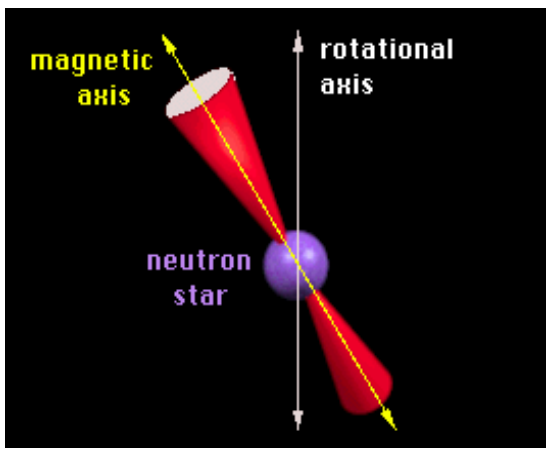

*Fig. 1: A schematic of a neutron star rotation with an inclined magnetic axis.*

Although this method boasts of the first definite discovery of an exoplanet, yet the total number of discoveries brought about by this method remain modest compared to the Doppler RV and transit techniques. The number of planets around pulsars should improve with major facilities such as Arecibo PALFA project and the soon-to-come up Square Kilometre Array (SKA) focussing on discovering many more pulsars in near future.

*Transit Method*

When a planet crosses in front of the stellar disk of its host star, there is a dip in the magnitude of the star due to the occultation by the planet in its orbital path. This dip repeats periodically indicating the orbital period. The amplitude of the dip will depend upon the area of the star blocked by the planet and hence indicates the size of the planet. A giant planet like Jupiter will, for instance, cause a dip of ~1% in the light curve of the host star whereas a terrestrial planet like earth will cause a dip of ~0.01%. It is absolutely essential to have a stable photometer with a very high S/N capability to measure dips with a precision better than $10^{-4}$. Moreover, monitoring the source over many orbits is essential to confirm the period. The Hubble Space Telescope time-series photometry[5] was used to observe four transits of the planet of star HD 209458.

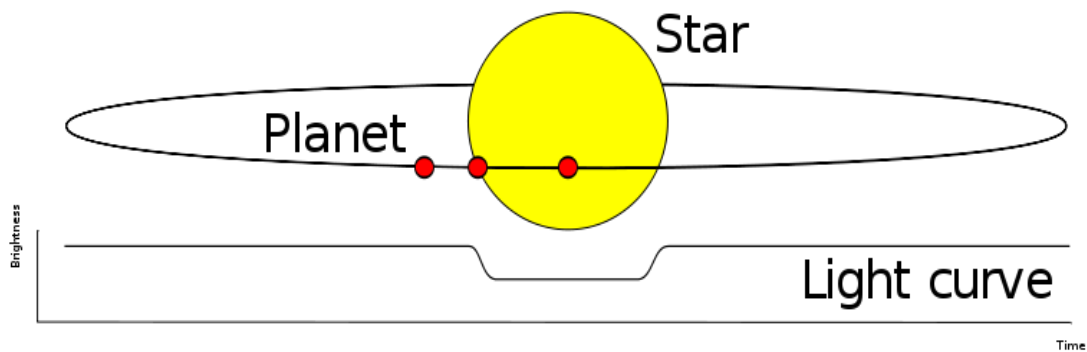

*Fig. 2: A planet transiting against the disk of its host star.*

This method however is dependent on the geometry of the system. The planetary transits are observable for planets whose orbits happen to be close to the line of sight from earth and the chance probability of such a thing happening is rather small. However this fact is compensated by looking for short period planets that are much easier to find. The main advantage of the transit method is that the size of the transiting body can be determined from the lightcurve. To prove that the body is a planet (mass less than ~0.01 $M_{sun}$), the results can be combined with the RV method to ascertain the mass, density of the planet, and hence learn something about the planet's physical structure. At the opposite side of the transit, the planet passes behind the host star and its heated face is eclipsed. This results in a very small sharp drop in the total light of the system, which in principle can be detected. This secondary eclipse can give an idea of the planet's temperature.

*Kepler*

It is a space observatory launched by NASA to discover Earth-like planets orbiting other stars. The spacecraft, launched in 2009, has been active for five years. The mission was specifically designed to survey a portion of our region of the Milky Way galaxy to discover Earth-size planets in or near the habitable zone and determine how many of the billions of stars in our galaxy have such planets. Kepler's only instrument is a photometer that continually monitors the brightness of over 145,000 main sequence stars in a fixed field of view. This data is transmitted to Earth and then analysed to detect periodic dimming caused by extrasolar planets that cross in front of their host stars. The Kepler space observatory is in a heliocentric orbit, so that Earth does not occult the stars, which are observed continuously, and so the photometer is not influenced by stray light from Earth. The photometer points to a field in the northern constellations of Cygnus, Lyra and Draco, which is well out of the ecliptic plane, so that sunlight never enters the photometer as the spacecraft orbits the Sun. As of July 2014, there are a total of 4234 candidates. Of these, 978 are confirmed exoplanets. Many planet candidates were found in the habitable zones of surveyed stars. In November 2013, a new mission plan named "K2" (also called "Second Light"), was presented for consideration, and on May 16, 2014, NASA has announced the approval. K2 would involve using Kepler's remaining capability, at a slightly reduced precision, to collect data for the study of supernova explosions, star formation and solar-system bodies such as asteroids and comets, and for finding and studying more exoplanets. In this revised mission plan, Kepler would search a much larger area in sky.

*Gravitational Microlensing*

Gravitational microlensing method for detecting exoplanets relies on the line of sight juxtaposition of two stars. According to Einstein's general theory of relativity, light bends due to gravity. Thus, if we have stellar light coming from a remote star and there is another star in the line of sight then the foreground star will act as a lens and will lead to amplification of the light coming from the background star. If there happens to be a planet going around the foreground lensing star, then there could be a brief intensification of light when the planet will also happen to be in the observer's line of sight along with the two stars. However, the amplification of the signal from the lensed star and planet lasts only for a limited time and cannot be predicted. Thus, the phenomenon is rare with a miniscule individual probability but monitoring of a dense field of millions of stars turns it into a finite possibility. Microlensing is well suited for finding low-mass planets and has successfully led to the detection of a few exoplanets.

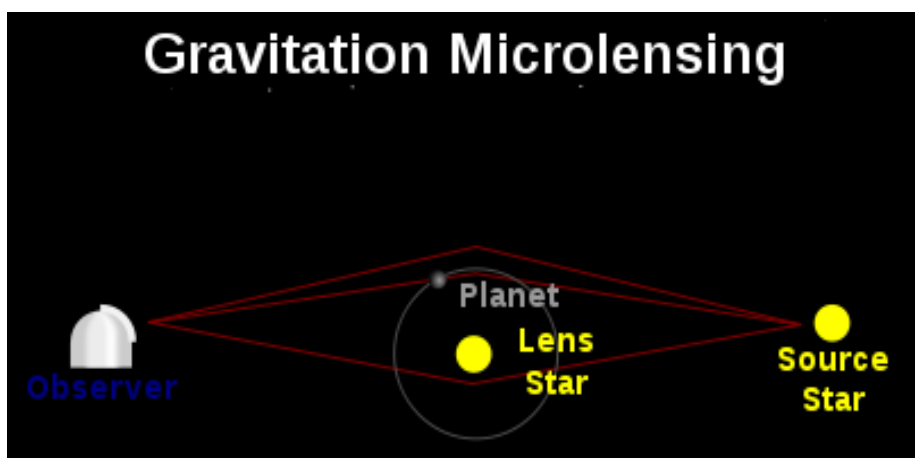

*Fig. 3: Gravitational microlensing by a star and the planet around it.*

*Astrometry*

This is one of the oldest methods for detecting a companion around the host star. The star's residual motion after subtracting its projected proper motion and the annual parallax is examined and if this motion shows a periodicity, it suggests the presence of a companion. The maximum orbital displacement $\alpha$ (in arcsec) of the star in the sky-plane, about the system's barycenter, will be given as,

$$\alpha = \frac{M_p}{M_*} \times \frac{a_p}{d}.$$

Here $M_p$ and $M_*$ are the masses of the planet and the star respectively, $a_p$ is the major axis of planet's orbit around the star in AU and *d* is the distance of the stellar system from Earth in parsecs. For a decade from 1963 to about 1973, there was a claim[6,7] of detection, by using astrometry, of a perturbation in the proper motion of Barnard's Star consistent with its having one or more planets comparable in mass with Jupiter. However later work showed[8] that changes in the astrometric field of various stars were correlated to the timing of adjustments and modifications that had been carried out on the refractor telescope's objective lens; the planetary discovery was an artefact of maintenance and upgrade work. Nevertheless, the upcoming European space Agency's Gaia mission's microarcsec precision ability to measure shifts in stellar positions should enable discoveries of exoplanets by this method. This technique would be more useful to look for planets across stars with wider orbits as against the RV technique, which is more suitable to look for shorter orbital period planets.

*Imaging*

Imaging the planet essentially means detecting the light intercepted and reflected by the planet from its host star. The amount of the reflected light by the planet depends on its distance from the star, its physical size as well as the atmosphere and nature of the surface (terrestrial/gaseous). What is important here is the brightness ratio of planet to star which for Sun-Earth system is about $10^{-10}$ in the visible region and it improves to about $10^{-6}$ in the infrared which makes IR direct imaging of exoplanets a possibility. This method works best for young planets that emit infrared light and are far from the glare of the star. With upcoming facilities such as Near Infrared Camera (NIRCam) and Mid-Infrared Instrument (MIRI) aboard the James Webb Space Telescope (JWST), imaging and characterizing exoplanets this way would become relatively easier.

*Radial Velocity*

The basic principle behind the ingenious method of radial velocity (RV) comes from the simple physics of the Doppler effect. Any rest-frame photon of wavelength $\lambda_0$ will be detected at a different wavelength $\lambda$ by an observer depending on whether the object is moving toward the observer (blue shift) or away from the observer (red shift), with λ given by,

$$\lambda = \lambda_0 \frac{1+\frac{v}{c}\cos(\theta)}{\sqrt{1-\frac{v^2}{c^2}}} \approx \lambda_0 \left[1+\frac{v}{c}\cos(\theta)\right]$$

Here *v* is the velocity of the source moving at an angle θ with respect to the direction of the source as seen in observer's reference frame, *c* is the velocity of light in vacuum (the approximation is valid for *v/c* <<1, applicable for planetary motions around their host stars). In the RV measurements of a star as a function of time, any periodic variations indicate the presence of a secondary around the primary. This method was brought to limelight with its first remarkable result of detection of a planet surrounding a main sequence star 51 Pegasi in 1995. Obtaining precise orbital parameters for super-Jupiters to Earth-like planets is possible by the RV measurements.

The RV technique has improved over the years. In 1953, radial velocities with a typical precision of 750 m/sec were used for cataloguing stars. With the need for detecting earth like planets in habitable zone of the host star requires precision of the order of a few cm/sec. For example a Jupiter like planet induces a RV shift of 12 m/sec on Sun whereas Earth at 1 AU causes RV shift of 9 cm/sec. Thus detection of smaller masses at distant orbits requires better precisions and stabilities. The simultaneous reference technique has been successfully employed in many spectrographs at different observatories around the world. With spectrographs such as SOPHIE, HARPS and India's indigenous PARAS installed at PRL having achieved precisions at the level of 1-2 m/sec (for HARPS below 1 m/sec) we are indeed closer to discovering more exoplanets in the habitable zone of the host stars.

Out of a total 1822 confirmed planets in 1137 planetary systems including 467 multiple planetary systems (as of 12 September 2014), about one-third of them are by radial velocity technique using high-resolution spectroscopy. The following table shows the number of exoplanets detected by various methods.

| *Method for detection* | *Number of exoplanets detected* |
|---|---|
| Radial Velocity | 574 |
| Transiting planets | 1147 (Confirmed by alternate methods as well) |
| Gravitational Microlensing | 32 |
| Timing | 15 |
| Direct Imaging | 51 |
| Transit time variation (TTV) | 3 |

*Table 1: Number of exoplanets detected by various detection techniques*

**Habitability**

Currently Earth is the only planet that we know which supports life and we still are in the hunt for similar other planets that could support life. For a habitable planet we essentially mean a planet which lies within a region around the star that could support life (of the type that we know); the region popularly known as the Goldilocks zone. Since we know of life only as it exists on Earth, we can at present imagine only similar type of conditions that are conducive to the formation and survival of life on Earth. Therefore, the habitable zone is calculated based on the position of Earth in the solar system and the amount of radiant energy it receives from Sun. This region has the right temperature and right pressure to support liquid water at the surface and thereby serves a possible place to harbour life in the current known form. Not only that even the orbit of the planet should be quite stable around the star to avoid too wild changes in the star-light received and the changes in the planet's atmosphere. Therefore, a planet in a binary star system may not be a good candidate for a habitable place as its orbit may not be very regular and the resulting fluctuations in the temperature and atmospheric conditions on its surface may not be very conducive to the survival of life on it. But there are counter arguments to this in the literature and it has been argued that such a system might in fact be more likely to lead to a start of life. Potentially, life could exist even more in binary systems than it does in single systems and indeed it could be possible that BOTH Alpha Cen A and B have planets conducive to life.

The phenomenon of a planet lying in the habitable zone of the host star depends on the orbital separation of the planet, the mass of the host star itself as well as the radiative flux emitted by the star. This habitable zone is also split into two regions, one being the conservative habitable zone where less massive planets like Earth and Venus can remain habitable and the extended habitable zone where super-Earth planets with stronger greenhouse effects could sustain liquid water on the surface.

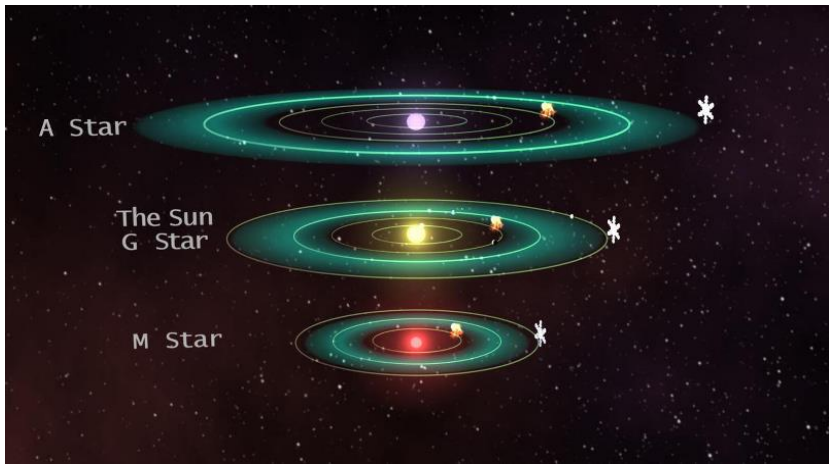

*Fig. 4: The exact location of a planetary system's habitable zone, a region with temperatures amenable to liquid water and life, depends on the type of host star. Credit: NASA*

Kepler Mission has found over 60 planets that could be lying in the habitable zone of their host stars. Kepler-186f situated 490 light years from Sun, ~10 % larger than Earth and rocky in nature, lies in the habitable zone. There are many similar planets like Kepler-22b, Tau Ceti e and f which are orbiting Sun-like stars, and some other planets like the super-Earth Gliese-667c as well as Gliese-581d which orbit around their host red dwarfs; all of these lie in the respective habitable zones of their host stars. Then there are Kepler-47b and 47c, the first transiting circumbinary system — one or more planets orbiting binary stars, and one of those planets is in the binary system's habitable zone (where liquid water may exist)! Another discovery is a five-planet system around a star called Kepler-62, some 1200 light-years away in the constellation Lyra. All five planets transit the star meaning their orbits appear to cross in front of their star as viewed from the Earth's perspective. Their inclinations relative to Earth's line of sight vary by less than one degree. This allows direct measurements of the planets' periods and relative diameters (compared to the host star) by monitoring each planet's transit of the star. The inner three worlds seem too hot for life, but planets Kepler-62e and Kepler-62f may be far more accommodating. They are 1.6 and 1.4 times the diameter of Earth respectively, and their orbits are within the boundaries of the habitable zone in which liquid water could exist. In fact another star system has been identified with up to seven planets – three of which could potentially host life – 22 light-years away. The likelihood that conditions could support life on at least one of those planets, given that there are three terrestrial-mass planets in the habitable zone of one system, is tremendous. It seems highly likely that **we are not alone**, so there is no reason for eremiphobia – fear of loneliness!

# The Possibility of finding intelligent life elsewhere in the Milky Way

There could be over 100 million planets in the Milky Way galaxy that could support complex life, though this does not necessarily imply that complex life does exist on that many planets. What it means that there are planetary conditions that could support life. Origin-of-life questions are not even addressed – only the conditions to support life. Complex life doesn't mean intelligent life – though it doesn't rule it out – but simply that organisms larger and more complex than microbes could exist in multitude forms. Using more than 1,000 planets and using a formula that considers planet density, temperature, substrate (liquid, solid or gas), chemistry, distance from its central star and age, something called the Biological Complexity Index (BCI) can be computed. The BCI calculations reveal that 1 to 2 percent of the planets could have a BCI rating higher than Europa, a moon of Jupiter thought to have a subsurface global ocean that could harbour forms of life.

Despite the large number of planets that could harbour complex life, the Milky Way galaxy is so vast that planets that support life could be very far apart. Then a question arises – Can we ever have a physical 'contact' with them or at least achieve some far-distance communication? At this stage it might be extremely difficult or perhaps almost impossible to state with any amount of certainty the probability of us contacting life elsewhere. Drake expressed this in the form of an equation for calculating the probability of detection of intelligent life within our Milky Way galaxy by an electromagnetic communication.

*Drake Equation*

Drake equation estimates the number of detectable extra-terrestrial (E.T.) civilizations in our Milky Way galaxy. The equation states,

$N = N_s \times f_p \times n_e \times f_l \times f_i \times f_c \times f_L$

$N$ = number of civilizations detectable in the Milky Way;
$N_s$ = estimated number of stars in the Milky Way;
$f_p$ = fraction of the stars having planets in orbits around them;
$n_e$ = average number of planets around a star with potential to host life as we know it;
$f_l$ = fraction of such planets that actually develop life;
$f_i$ = fraction of the planets that actually develop intelligence on human level;
$f_c$ = fraction of civilizations that develop electromagnetic radiation emitting technologies;
$f_L$ = fraction of civilizations at any time emitting electromagnetic signals in space.

The equation was written in 1961, with every factor very uncertain. The only change that has happened in the intervening years is that $f_p$, the fraction of stars having planets orbiting them, is perhaps now close to one, although the remaining factors still remain almost as uncertain as ever. Of course there could be other factors like the average time scale an intelligent civilization may last. Based on Kepler space mission data, it is estimated there could be billions of Earth-sized planets orbiting in the habitable zones of sun-like stars and red-dwarf stars within the Milky Way galaxy, with the next nearest planet in the habitable zone being as little as 12 light-years away from Earth. However the number that actually harbours life could be much smaller.

If we drop the word "intelligence" and the ensuing factors from the Drake equation and concentrate on only the simplest forms of life, then their occurrence probability may increase by a large value; however, their detectability from Earth becomes much less probable. First and foremost, we cannot travel up to them as it has been calculated and shown that a return trip to a planet merely 12 light-years away is almost impossible, even using our most-efficient controlled thermonuclear reactions, as it might need fuel mass equal to one-third of our Galaxy[16]. Thus, it does not seem possible that we will ever be able to touch fingers with any E.T. or have a miraculous transformation from an embracing (embarrassing to most conformist zealots the mere idea itself!) encounter with a Jadoo from beyond our own Solar system. Of course the same energy constraints will be applicable to any 'guest from yonder' planning to pay us a visit (thus ruling out any real UFOs)!

Therefore any communication has to be through an electromagnetic media like radio waves. It is also more likely that we will be successful in making such a "contact" if we try to 'listen' to them instead of a more active transmission on our part. The argument goes something like this: Our modern science is only about 400 years old, and our radio communication skills are less than a century old. The alien intelligence that we might ever get in touch with will most likely be much more advanced than us in their communication techniques and will be able to transmit much higher levels of electromagnetic power and thereby able to reach much farther than we can (if they are not so advanced then it is very unlikely that we will be able to communicate with them). Therefore it is much more sensible to try to 'listen' to them.

Instead of finding aliens with radio technology, Seager[10] has revised the Drake equation to focus on simply the presence of any basic form of life. Her equation can be used to estimate how many planets with detectable signs of life might be discovered in the coming years. Focusing on M stars, the most common stars our neighbourhood that are smaller and less luminous than our Sun, her calculations suggest that two inhabited planets could reasonably turn up during the next decade.

Of course there exists an additional complication that some of the planets in the habitable zone, especially around the M dwarf stars, might be tidally locked to their host star which might make them incapable of sustaining life due to their temperature extremes and other harsh conditions on the surface of such planets[17]. And even if planets are in the habitable zone, the number of planets with the right proportion of elements is however, difficult to estimate, with numerous gas giants in close orbit. The possibility of life on exomoons of gas giants (such as Jupiter's moon Europa, or Saturn's moon Titan in our solar system) adds further dimensions to this enigmatic problem.

**Exomoons**

On the lines of current hunt for signatures of life across the planets in outer solar systems, a good thought could be raised on probing the less explored territory of exomoons, a natural satellite to an exoplanet, and the speculation of habitable conditions at such moons. Looking at our own backyard, we have Jupiter and Saturn hosting more than 60 moons around their orbits. The potential of the moons, Europa orbiting Jupiter and Titan orbiting Saturn, on possibilities of sustaining life, naturally opens gateways to postulate the presence of similar exomoons in systems apart from ours. The presence of such an exomoon might also lead to greater probability of finding life across the host planet. We do not yet know for sure whether the presence of moon across Earth has really boosted the occurrence of life or it just happens to be a mere coincidence. Further, a Jupiter-like gaseous exoplanet may have moons with solid rocky surface for higher forms of life to evolve on. Currently with no exomoon detection so far, we are not in an undaunted position to make any claims but this does act as a big motivation to look for such moons apart from ours.

The theory of exomoon detection proposes the use of transit timing. The current theories of planetary formation mechanism are vaguely understood. Our current understanding of formation of planets by core accretion model or fragmentation of the dust cloud is still dubious. The presence of natural satellites across such planets may lead to insights in their formation mechanism, stability across the planet and further evolution. Although the highlight of the search is looking for habitable exomoons, the achievable sensitivities with our current instrumentation will not allow us to probe the atmosphere of any exomoon. Thus, a mere detection as a first step will quench the preliminary hunt for this new class of extra-solar objects.

A systematic search to look for exomoons has begun with the planetary candidates found by Kepler. This project is termed as Hunt for Exomoons with Kepler (HEK). The transit time variation (TTV) and transit duration variation (TDV) techniques are used to characterize the orbit of exomoons. These techniques along with some other detection methods are briefly discussed below.

*Finding exomoons*

- Direct Imaging: A Jupiter sized planet as seen against Sun at a distance of 5 AU will be about a billion times fainter than the host star. A satellite being smaller than the host planet will be a couple of orders of magnitude further fainter than the planet. The Earth Moon distance from 10 pc will give an angular separation of 0.25 milliarcsec. The detection of such a natural satellite by the imaging method seems unlikely to materialise with the current imaging technologies.

- Radial Velocity: The sensitivity of Radial velocity method is reaching finer precision which is enabling the search for Earth-like planets in habitable zones. However the presence of a moon across a planet will reflect only in the amplitude of the wobble which is used to detect the mass of the orbiting body. The presence of a moon can thereby be disguised easily by an assumption of a heavier planet against the host star. In order to rightfully detect the presence of an exomoon around the planet, one will need sufficient light from the planet itself to enable radial velocity of the planet to detect the presence of a companion to its orbit. This will need hours of exposures on telescopes with larger diameters. Thus, detection of an exomoon by radial velocity method seems challenging within current scenarios.

- Transit time effects:
    1. Transit time variation (TTV): For a planet on a nearly edge-on orbit around the host star, we expect to observe the transit of the star by the planet. If this planet also harbours a satellite around it, then the planet will make a wobble across the planet-satellite barycenter. This will shift the transit occurrences to be periodically earlier and/or later. The reflex motion induced on the planet by the exomoon will lead to transit time variations of the order of several seconds to a few minutes. However, a major hurdle in the detection of exomoons by this method is that a variety of other effects like the presence of another planet or star, tidal distortions between two interacting bodies could induce similar variations in transit. Hence it is difficult to make sure whether the variation is caused only by the presence of an exomoon.
    2. Transit duration variation (TDV): Since, we expect the transit time to alter with the presence of an exomoon, we also expect the duration of this transit to change due to the change in velocity which is caused by the motion of the planet across the planet-satellite barycenter.

TTV is capable of detecting heavier moons at larger separations whereas TDV is efficient in the detection of closer moons. Thus, TDV serves a similar solution like the Radial Velocity method which assists photometry for detection of exoplanets with the only difference that in TDV we are concerned about the tangential velocity and not the radial velocity. TDV can be induced by parallax effects. However when these signals are combined with TTV signal, it will be 90° out of phase from TTV and thereby offer a unique exomoon signature.

## Conclusions

The nearest possible exoplanet, if confirmed, will be Alpha Centauri Bb orbiting Alpha Centauri B, which is one of three stars in a triple star system nearest to our Solar system. The Kepler mission has identified about 4000 planetary candidates, several of them being nearly Earth-sized and located in the habitable zone. Although, the question of habitability and the presence of life is very intriguing and still far from any possible solution, at least we are presently in a strong position to comment on the statistics of the exoplanets discovered till date. Every discovered exoplanet serves as a tool to characterise the planet formation mechanism, star-planet interaction and their further evolution. The presence of giant planets around host stars raises questions on planetary migrations or capture of the planet in the star's gravitational field. Moreover, with improvement in space based photometry missions like KEPLER (HEK), which are able to correctly model the transit to an accuracy of few seconds, detection of exomoons will not remain a distant dream for sure. Hopefully, a study of TTV and TDV together will lead to detection of exomoons. This will certainly shed more light on the theories of planet and satellite formation across stellar systems. Thus, we earthlings must keep on anticipating. Little we know, we might stumble upon some fascinating '*Pandora'* of *Avatar* fame!

**Further Reading:**

1) Wolszczan, A. & Frail, D., Nature, 355,145 (1992)
2) Mayor M. & Queloz D., Nature, 378, 355, (1995)
3) Butler, P. & Marcy, G., ApJL, 446, 153, (1996)
4) Marcy, G & Butler, P., ApJL, 464, 147, (1996)
5) Timothy, B., et al, ApJ, 552, 699 (2001)
6) Van de Kamp, P., AJ 68, 515 (1963)
7) Van de Kamp, P., Vistas in Astronomy 26, 2 (1982)
8) Hershey, J. L., AJ, 78 (6): 421 (1973)
9) Doyle, L., Extra-solar Planets: XVI Canary Islands Winter School of Astrophysics, Eds. Deeg, H., et al., Cambridge University Press (2007), p. 1
10) Seager, S., Exoplanets, University of Arizona Press (2010), p. 1
11) Marcy, G., Ann. Rev. Astron. Astrophys., 36, 57 (1998)
12) Kipping, D., et al., Astrophys. J., 750, 115 (2012)
13) Sartoretti, P. & Schneider, J., Astron. Astrophys. Suppl., 14, 550 (1999)
14) Sagan, C., Carl Sagan's Cosmic Connection – An Extra-terrestrial Perspective, Cambridge University Press (2000).
15) Purcell, E., Interstellar Communication, ed. Cameron, A. G. W., Benjamin Inc. (1963), p. 121
16) Singal, T. & Singal, A. K., Planex News Lett. 3, issue 1, 22 (2013)
17) Singal, A. K., Planex News Lett. 4, issue 2, 8 (2014)
18) Irwin, L., et al., Challenges, 5, 159 (2014) doi:10.3390/challe5010159
19) Barclay, T., et al., Astrophys. J., 768, 101 (2013) doi:10.1088/0004-637X/768/2/101
20) Borucki, W., et al., Science, 340, 587 (2013) doi:10.1126/science.1234702.
21) http://www.space.com/18790-habitable-exoplanets-catalog-photos.html
22) http://en.wikipedia.org/wiki/Exoplanet
23) http://www.astrobio.net/news-exclusive/habitable-binary-star-systems/

**Priyanka Chaturvedi,** *Astronomy & Astrophysics Division*, *Physical Research Laboratory, Ahmedabad, India.* **Email:** priyanka@prl.res.in

**Ashok K Singal,** *Astronomy & Astrophysics Division,,Physical Research Laboratory, Ahmedabad, India*. **Email:** asingal@prl.res.in